\documentstyle[epsfig,graphicx,prb,aps,twocolumn,floats]{revtex}

\begin{document}

\twocolumn[\hsize\textwidth\columnwidth\hsize\csname @twocolumnfalse\endcsname

\title{Calculation of the optical response of C$_{60}$ and Na$_{8}$
using time-dependent density functional theory and local orbitals}
\author{Argyrios Tsolakidis, Daniel S\'{a}nchez-Portal$^*$, and Richard M. Martin}
\address{Department of Physics and Materials Research Laboratory\\
University of Illinois at Urbana-Champaign, Urbana,
Illinois 61801}
\date{\today}
\maketitle

\begin{abstract}

We report on a general method for the calculation of the frequency-dependent
optical response of clusters based upon time-dependent density functional
theory (TDDFT). The implementation is done using explicit propagation in the
time domain and a self-consistent program that uses a linear combination of
atomic orbitals (LCAO).
Our actual calculations employ the SIESTA program, which is designed
to be fast and accurate for large clusters. We use the adiabatic local density
approximation to account for exchange and correlation
effects. Results are presented for the imaginary part of the linear
polarizability, $\Im \alpha(\omega)$, and the dipole strength function, $S(\omega)$, of C$_{60}$
and Na$_{8}$, compared to previous calculations and to experiment.
We also show how to calculate the integrated
frequency-dependent second order non-linear
polarizability for the case of a step function electric 
field, $\tilde{\gamma}_{step}(\omega)$,
and present results for C$_{60}$.
\end{abstract}
]

\newpage

\section{Introduction}

Although density functional theory (DFT)~\cite{Hohen,Kohn1} is a very
successful theory for the ground
state properties, the excited states calculated within the Kohn-Sham scheme
often are much less successful in describing the optical response and the
excitation spectra. The solution to this problem,
in principle, is
the extension of DFT to the time-dependent systems. It is
interesting to note that the first calculation~\cite{Zangwill} using
TDDFT preceded any formal development and it relied
heavily on the analogy with the time-dependent Hartree-Fock method. The first
steps towards the formulation of TDDFT were done by Deb and Gosh~\cite{Deb,Ghosh} who
focused on potentials periodic in time, and by Bartolotti~\cite{Bar1,Bar2} who focused on
adiabatic processes. Runge and Gross~\cite{Runge} established the
foundations of TDDFT for a generic form of the time-dependent potential. TDDFT
was further developed~\cite{Kohn2,Kohn3} to acquire a structure that is very similar to
that of the conventional DFT. A very interesting feature of TDDFT, that does
not appear in DFT, is the dependence of the density functionals on the initial
state. For more information about TDDFT the reader is advised to read the
authoritative reviews of Gross, Ullrich, and Gossmann,~\cite{Gro1} and Gross,
Dobson, and Petersilka.~\cite{Gro2}

The polarizability describes the distortion of the charge cloud caused by
the application of an external field. It is one of the most important response
functions because it is directly related to electron-electron interactions,
and correlations. In addition, it determines the response to charged
particles, and optical properties. A quantity of particular interest is the
dipole strength function, $S(\omega)$, which is directly related to the
frequency-dependent linear polarizability, $\alpha (\omega)$, by
\begin{equation}
\alpha(\omega)=\frac{e^{2} \hbar}{ m } \int_{0}^{\infty} \frac{S(\omega^{\prime})
d\omega^{\prime}} { {\omega^{\prime}}^{2}-{\omega}^{2}}. \label{eq-first}
\end{equation}
By taking the imaginary part of Eq. (\ref{eq-first}) we obtain
\begin{equation}
S(\omega)=\frac{2 m}{\pi e^{2} \hbar} \omega \Im \alpha(\omega).
\end{equation}
The dipole strength function, $S(\omega)$, is proportional to the
photoabsorption cross section, $\sigma(\omega)$, measured by most experiments
and, therefore, allows direct comparison with experiment. In addition, the integration of
$S$ over energy gives the number of electrons, $N_{e}$, ({\it f-sum rule }) i.e.
\begin{equation}
\int_{0}^{\infty} dE S(E) = \sum_{i} f_{i} = N_{e},
\end{equation}
where $f_{i}$ are the oscillator strengths. This sum rule is very important because it provides an
internal consistency test for the calculations, indicating
the completeness and adequacy of the basis set used for the
computation of the optical response.

Optical probes are some of the most successful experimental tools that allow
access to the properties
of clusters. Consequently, there are
many calculations of the optical
response of small atomic aggregates.
In particular, there exist several theoretical
studies of
the examples chosen here, C$_{60}$ and
Na$_{8}$. This allows us to calibrate the accuracy of our
method in comparison with other computational schemes.
One of the first {\it ab initio} calculations of the
dipole response
of atomic clusters within TDDFT was performed by Yabana and Bertsch,~\cite{Yabana} who studied large
sodium and lithium clusters, and the
C$_{60}$ molecule
using a real time and space approach. For small Na clusters,
Vasiliev~{\it et al.}~\cite{Vasiliev} calculated the absorption cross section using the
time-dependent density functional response
theory (TD-DFRT) developed by Casida.~\cite{Casida}

The purpose of this work is to propose a method that will have significant
advantages for the calculation of the polarizability of large clusters,
reducing considerably the
computation time while retaining the desired
accuracy. This paper is organized as follows: In section
\ref{sec-Method}, we describe the method of calculation. In section
\ref{sec-Solut}, we give details about the way we solve the time-dependent
Kohn-Sham equation
and briefly summarize other methods available. In section \ref{sec-Results}, we
present an overview of relevant calculations and the results of our
calculation for C$_{60}$ and Na$_{8}$. We compare our results with other calculations and
experiments. In section \ref{sec-Non}, we describe the calculation and present
the results for the
imaginary part of the integrated frequency-dependent second order non-linear
polarizability for the case of a step function electric field, 
$\Im \tilde{\gamma}_{step} (\omega)$, for C$_{60}$.
In section \ref{sec-Concl}, we give the conclusions.

\section{Method of Calculation}\label{sec-Method}

\subsection{Electronic structure calculations}

Our method involves the
description of the electronic states using
linear combination of atomic orbitals (LCAO). Because the size of the LCAO
basis is small, compared with other usual choices like plane waves or real
space grids, the TDDFT calculations can be done
efficiently using the techniques described
below. Our scheme is based on the
SIESTA~\cite{Daniel,Artacho,Ordejon} code, which is used to compute the initial wavefunctions and the
Hamiltonian matrix for each time step.
SIESTA is a general-purpose DFT code which uses a local basis,
and has been
specially optimized to deal with large systems. As such, it represents an
ideal tool for treating large clusters. Core electrons are replaced by
norm-conserving
pseudopotentials~\cite{TM}
in the fully nonlocal
Kleinman-Bylander~\cite{Kleinman}
form, and the
basis set is a general and flexible linear combination of
numerical atomic orbitals (NAOs), constructed from the
eigenstates of the atomic pseudopotentials.~\cite{Artacho,Sankey}
The NAOs are confined, being
strictly zero beyond
a certain radius.
In addition, the electron wavefunctions and
density are projected onto a real space grid in order to calculate the Hartree
and exchange-correlation potentials and their matrix elements.

The use of confined NAOs is very important
for the efficiency of the SIESTA code.
With them, by
exploiting the explicit sparseness of the
Hamiltonian and density matrices,
the computational cost for the
construction and storage of the Hamiltonian
and the electronic density can be
made to scale
linearly with the number of atoms, in the limit of large systems.
Therefore, a considerable effort
has been devoted to obtain orbital bases that would meet
the standards of precision of conventional first-principles
calculations, while keeping their range as small as possible.
A simple scheme
for the generation of transferable bases that satisfy both
requirements was presented in Refs.~\onlinecite{Artacho} and \onlinecite{Junquera}.
These bases, which we utilize in
this work,
have been successfully applied to study the ground state properties
of very different systems, ranging from insulators to metals, and
from bulk to surfaces and nanostructures.~\cite{Ordejon}
It is not obvious however that these confined basis sets
will be
also adequate for the TDDFT calculation of the optical response.
In this paper
we show that, at least for the two systems considered,
the optical absorption can be accurately calculated
using basis of NAOs with reasonable confinement radii, and
a moderate number of orbitals per atom.
Our results are
in good agreement with other TDDFT calculations using computationally
more demanding basis sets.

Our approach is to carry out the calculations in the time domain, explicitly
evolving the wavefunctions. We consider a bounded system in a finite electric
field, i.e. the Hamiltonian includes a perturbation ${\Delta} H=-{\bf E} \cdot
{\bf x}$. For the linear response calculations in this paper we have
set the value of this field to
0.01~eV/\AA.
The system is
solved
for the ground state using standard time
independent density functional theory.~\cite{Efield}
Then we switch off the electric
field at time $t=0$, and for every subsequent time
step we propagate the occupied Kohn-Sham eigenstates
by solving the time-dependent Kohn-Sham equation ($\hbar=1$)
\begin{equation}
i \frac {\partial \Psi} {\partial t} = H \Psi, \label{eq-TDKSE}
\end{equation}
where H is the time-dependent Hamiltonian given by
\begin{equation}
H = - \frac {1} {2} \nabla^{2} + V_{ext}({\bf r},t) +
\int \frac { \rho({\bf r}^{\prime},t)}{|{\bf r}-{\bf r}^{\prime}|}
d{\bf r}^{\prime} + V_{xc}[\rho]({\bf r},t). \label{eq-H}
\end{equation}
The calculation of the exchange-correlation potential is done using the
adiabatic local density approximation (ALDA) where $V_{xc}$ takes the form
\begin{equation}
V_{xc}[\rho]({\bf r},t) \cong \frac {\delta E_{xc}^{LDA}[\rho_{t}]}{\delta
\rho_{t}(\bf r)} =  V_{xc}^{LDA}[\rho_{t}]({\bf r}).\label{eq-xc}
\end{equation}
$E_{xc}^{LDA}[\rho_{t}]$ is the exchange-correlation energy of the
homogeneous electron
gas.~\cite{PZ}
It is important to notice that the $V_{xc}$ in ALDA
is local both in time and space. For every time step we solve
Eq. (\ref{eq-TDKSE}), and from the new wavefunctions
we construct the new density matrix
\begin{equation}
\rho^{\mu \nu}(t) = \sum_{i\; occ} c_i^{\mu}(t) c_i^{\nu}(t),
\label{eq-den-mat}
\end{equation}
where $c_i^{\mu}(t)$ are the coefficients of the occupied
wavefunctions which correspond to the basis orbitals $\phi_{\mu}({\bf r})$.
$\rho^{\mu \nu}(t)$ has to be calculated and stored for
overlapping orbitals only.
The electron density is then obtained by
\begin{equation}
\rho({\bf r},t)= \sum_{\mu, \nu} \rho^{\mu \nu}(t) \; \phi_{\mu}({\bf r})
\phi_{\nu}({\bf r}), \label{eq-den}
\end{equation}
and used for the calculation of the Hamiltonian in the new cycle.

\subsection{Calculation of the polarizabilities}

For every time step we calculate the dipole moment ${\bf D}(t)$ of the electrons
in the cluster.
This defines the response to all orders and
the frequency dependent response is found by the Fourier transform
\begin{equation}
{\bf D}(\omega) \equiv \int dt e^{i\omega t - \delta t} {\bf D}(t).
\label{eq-fourier}
\end{equation}
In our case we Fourier transform the dipole moment only for t$>$0.
It is necessary to include a damping factor $\delta$ in order to perform
the Fourier transform. This damping factor gives the minimum width
of the peaks of the imaginary part of the response.
Physically, it can be regarded as an approximate way to account for
broadening.
To linear order
the polarizability is given by $ {\bf D} (\omega)= \alpha(\omega) {\bf E}(\omega)$,
so that
\begin{equation}
\Im \alpha(\omega) = \omega \frac{ \Re D(\omega)}{E},
\label{eq-lin0}
\end{equation}
where the field is given by $E(t) = E \; \theta(-t)$. After Fourier
transforming the dipole moment we obtain the
elements of the frequency-dependent polarizability
tensor $\alpha_{ij}(\omega)$.
We repeat the calculation
with the electric field along different axis unless the symmetry is high
enough that this is not needed.
The average linear polarizability is given by
\begin{equation}
< \alpha(\omega) > = \frac {1} {3 } Tr \{\alpha_{ij}(\omega)\}.\label{eq-pol}
\end{equation}
The choice of the coordinate system does not affect the average polarizability
because of the rotational invariance of the trace.

\subsection{Solution of the time-dependent Kohn-Sham equation}\label{sec-Solut}

Efficient solution of the time-dependent
Kohn-Sham equation (Eq. (\ref{eq-TDKSE})) is of particular
interest because together with the calculation of the Hamiltonian, they are
the most time consuming parts of the calculation. In this section we describe our
approach of solving   Eq. (\ref{eq-TDKSE}),
as well as other existing methods used for the same purpose.

In the LCAO formalism Eq. (\ref{eq-TDKSE}) takes the form
\begin{equation}
i \frac {\partial c} {\partial t} = S^{-1} H c\label{eq-LCAO}
\end{equation}
where $S$ is the overlap matrix between the orbitals and $c$ is the column of the
coefficients of the local orbitals.
The overlap matrix is fixed for a given atomic configuration, hence we have
to calculate and invert it only once.

The formal solution of Eq. (\ref{eq-LCAO}) is
\begin{equation}
c(t)=U(t,0)c(0)=T exp\left( - i \int_{0}^{t} S^{-1}H(t^{\prime})
dt^{\prime}\right) c(0), \label{eq-sol}
\end{equation}
where $T$ is the time ordering operator. The most elementary solution is
obtained by breaking the total evolution operator into evolution operators of
small time durations
\begin{equation}
U(t,0) \simeq \prod_{n=0}^{N-1} U((n+1)\Delta t, n \Delta t),\label{eq-appr}
\end{equation}
where $\Delta t = \frac{T_{tot}}{N}$ and
\begin{equation}
U(t + \Delta t, t)= exp \left( - i S^{-1} H(t) \Delta t \right).
\label{eq-appru}
\end{equation}
$T_{tot}$ is the total time that we allow the system to evolve.
The differences among propagation schemes arise from the way the exponential
in Eq. (\ref{eq-appru}) is approximated.
In our approach, we approximate the exponential in Eq. (\ref{eq-appru}) with
the Crank-Nicholson operator.~\cite{Varga} The coefficients
between the steps $n+1$ and $n$ are related by the equation
\begin{equation}
c^{n+1}= \frac {1-i S^{-1}H(t_n)
\frac{ \Delta t }{2}}{1+i S^{-1}H(t_n) \frac{ \Delta t}{2}}
\; \; c^{n}. \label{eq-cn}
\end{equation}
This method is unitary, strictly preserving the orthonormality of the
states for an arbitrary time evolution.
For time independent Hamiltonians it is also explicitly time reversal
invariant, and exactly conserves energy.
In practice, with a suitable choice of $\Delta t$, the
energy is satisfactorily conserved even when the Hamiltonian changes with
time. For example, in the calculations described below, the drift of the
total energy at the end of the simulation ($\sim$130~fs in both cases) was only
$\Delta E_{tot}/E_{tot}\sim 3 \times 10^{-7}$ for C$_{60}$ and,
$ \sim 8 \times 10^{-6}$ for Na$_8$, after N$_{C_{60}} \sim 6100$ and
N$_{Na_8} \sim 2800$ time steps, respectively. The larger energy drift in the
case of Na$_{8}$ is attributed to the use of larger time step.
The method is stable when $\Delta t \Delta E_{max}<< 1$,
where $\Delta E_{max}$ is the range of the eigenstates of $S^{-1}H$.
We can increase the
stability of the solution if we include more terms of the expansion in the
numerator and denominator of the Crank-Nicholson operator, i.e.
\begin{equation}
c^{n+1}= \frac {
1-i S^{-1}H \frac { \Delta t }{2}  - \frac {1} {2}(S^{-1}H
\frac { \Delta t }{2})^{2} +i \frac {1}{6}(S^{-1}H \frac {\Delta t }{2})^{3} }
{1+i S^{-1}H \frac {\Delta t}{2} - \frac {1} {2}(S^{-1}H \frac
{\Delta t }{2})^{2}  -i \frac{1}{6}  (S^{-1}H \frac { \Delta t }{2})^{3} } \; \;
c^{n}. \label{eq-cn2}
\end{equation}
By including more terms in the expansion it is possible either to increase
the time step
preserving the accuracy,
or to increase the accuracy of the dynamics and the energy conservation for a
given time step.
The main
advantage of using a bigger time step is
the saving of time
because we have to
calculate the Hamiltonian fewer times.
The energy resolution will not be
affected since it depends on the total time that we allow the system to
evolve.

The method presented in this work has many similarities with that described
by
Yabana and Bertsch,~\cite{Yabana} the main difference
being the use of an LCAO basis set in the present case.
However, this is a key difference because
the size of the matrices used in the
calculations is considerably smaller compared to other basis choices.
In addition, our method has other advantages
associated with the real time formulation of TDDFT.
Only occupied states are used in the calculation, in contrast to the
perturbative approach~\cite{Casida,Ehren} where there is a sum over the
excited
states of the system. The implementation is relatively simple, since we use
essentially the same
operations as already used
to find the ground state properties. It is also advantageous that nonlinear
effects can be included in a
straightforward way. One disadvantage of the real time approach is the
calculation of the
Hamiltonian for every time step. Although this is not an attractive feature
there is no other way to calculate the time evolution of the system.

There are many other ways to approach the solution of the equations.
For completeness, we discuss in the Appendix several methods that could be
of potential relevance for our calculations.  One of
the main goals of these methods is to enable longer time
steps.  This would be a great advantage in our work; however, the
fact that the self-consistent Hamiltonian changes as a function
of time, limits their use.

\section{Discussion of Results}\label{sec-Results}

\subsection{Small metal clusters: Na$_{8}$}

The first calculation we performed is the optical response of
Na$_{8}$. The main purpose of this calculation was to investigate the accuracy
of our method
in the case of a small cluster, where the effects related to the confinement of
the orbitals should be more noticeable
and where the size of our basis is much smaller than in previous
calculations using real space grids.~\cite{Vasiliev}
Na$_{8}$
is the smallest
closed shell Na cluster that its optical response exhibits the presence of a
plasmon which is experimentally observed at 2.53 eV.~\cite{Selby,Wang} The width of the
plasmon is due to Landau damping.~\cite{Ekardt}

Previous work can be grouped into two types: earlier work on
jellium spheres~\cite{Selby,Ekardt,Selby2,Yannouleas}
that reproduces the qualitative features but not
the quantitative energies of the peaks, and more recent
work~\cite{Vasiliev,Bouna} that takes into account the detailed atomic structure and
is in general in very good agreement with experiment.~\cite{Selby,Wang}
In the first category are the calculations of
Selby ${\it et~al.}$,~\cite{Selby,Selby2} who calculated the
photoabsorption cross section using the modified Mie theory, which is
a classical theory. The plasmon was
found
to be
at $\sim$ 2.76 eV.
By using the self-consistent jellium
model in the time dependent local density approximation (TDLDA),
first introduced by Ekardt,~\cite{Ekardt} Yannouleas
{\it et al.}~\cite{Yannouleas} calculated  the photoabsorption cross
section of Na$_{8}$ and predicted the plasmon position at 2.82 eV.

\begin{figure}[t]
\centering
\epsfig{file=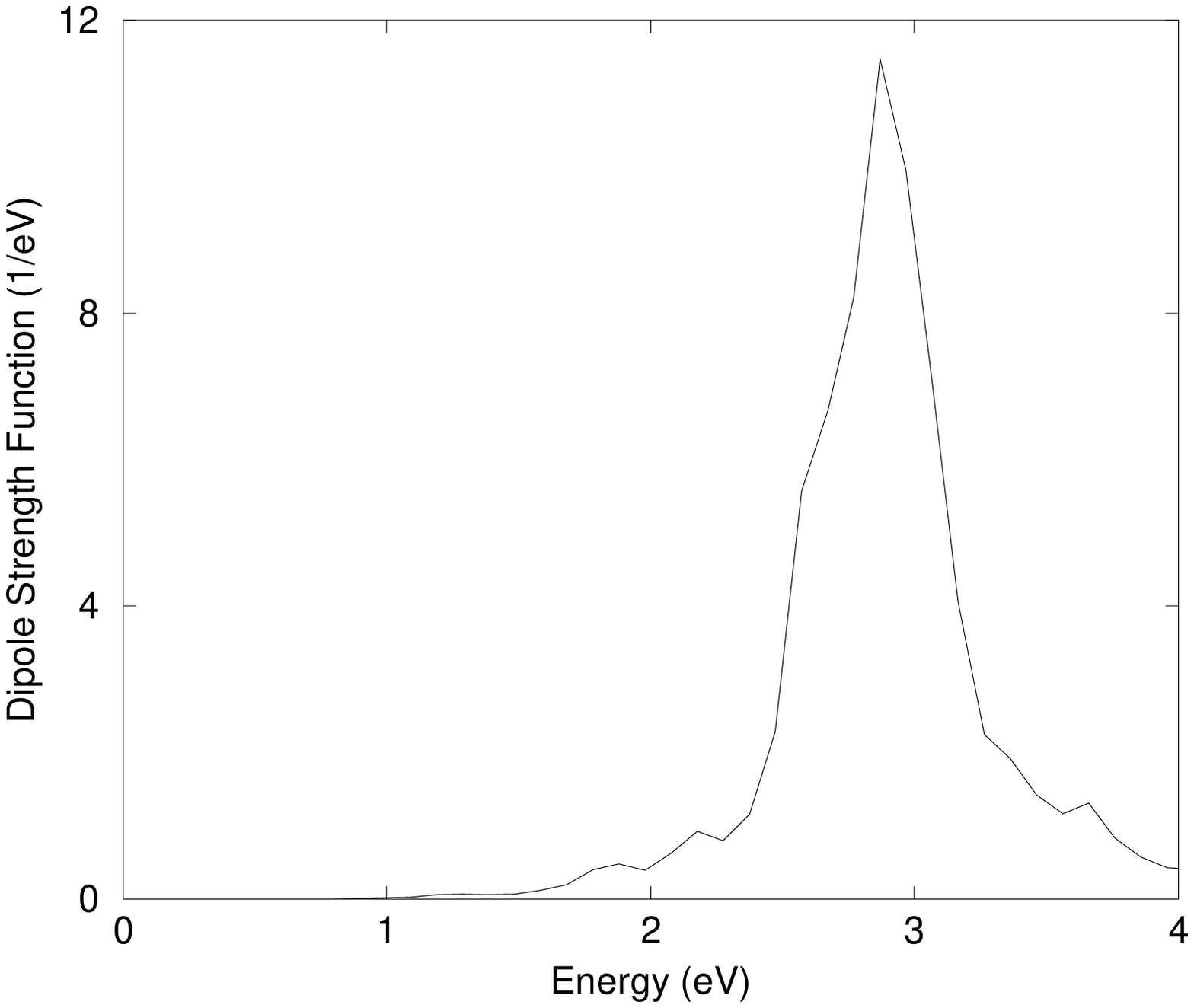, width= 8 cm, angle = 0}
\caption{Dipole strength function of Na$_{8}$ vs. energy.}
\vspace{0.5 cm}

\centering
\epsfig{file=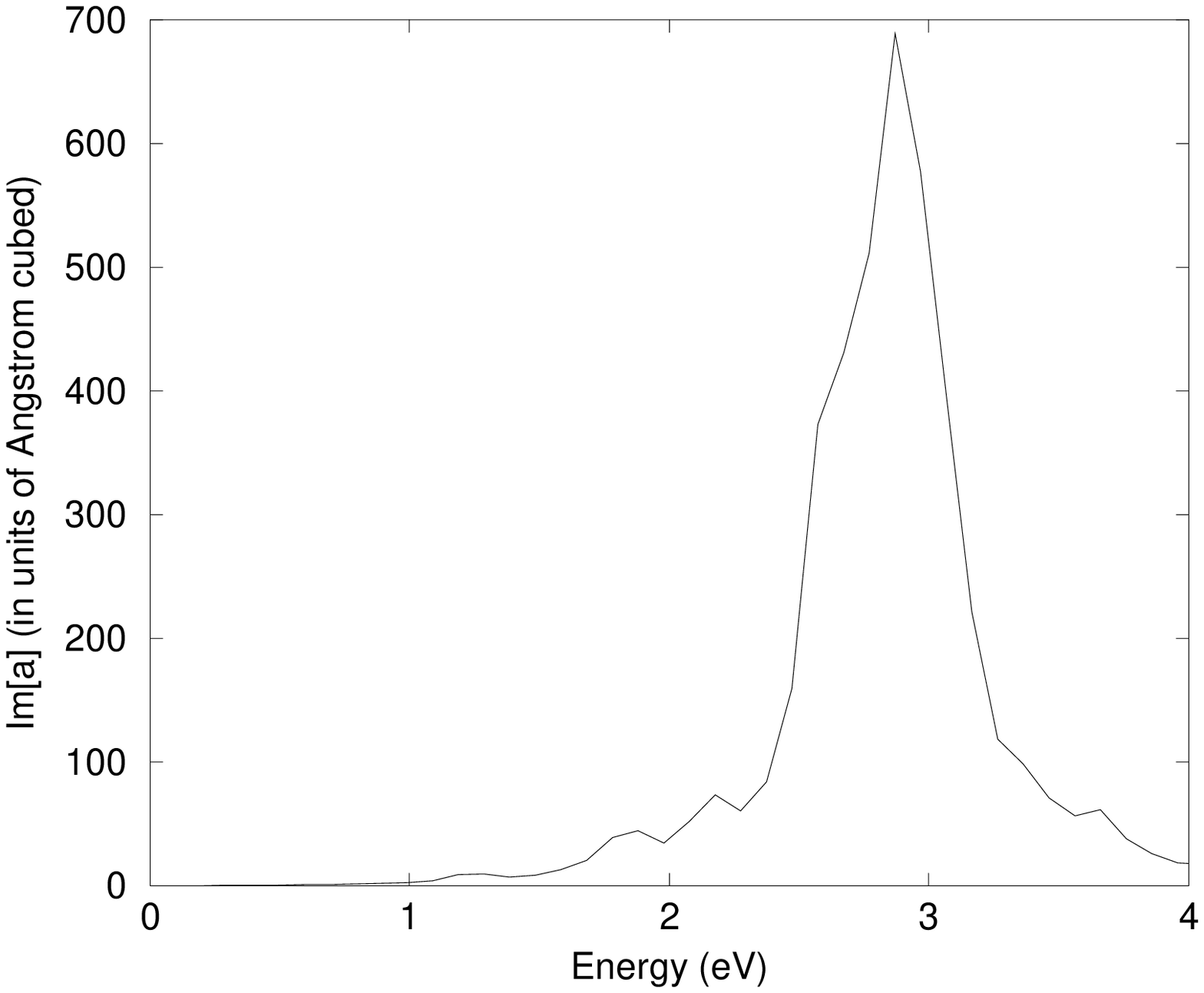, width= 8 cm, angle = 0}
\caption{Imaginary part of the linear polarizability of Na$_{8}$
   vs. energy.}
\end{figure}

Bouna\v{c}i\'{c}-Kouteck\'{y} {\it et al.}~\cite{Bouna} calculated the absorption
spectrum of  Na$_{8}$ using
the configuration-interaction method (CI).
Because the all-electron calculation is
computationally very
demanding, they obtained the excited states by a non-empirical effective core
potential
corrected for the core-valence correlation using a core
polarization potential.
The position of the plasmon was predicted at
$\sim$ 2.55 eV. Vasiliev {\it et al.}~\cite{Vasiliev} calculated the
photoabsorption cross
section using TD-DFRT.~\cite{Casida}
Their calculations made use of norm-conserving pseudopotentials
and a real-space mesh as a basis set.
The position of the plasmon agreed
with the photoabsorption experiments of  Selby {\it et al.}~\cite{Selby} and
Wang {\it et al.}~\cite{Wang} within 0.1-0.2 eV.

In our calculation we
let the system evolve for the total time of T=31.42~eV$^{-1}$.
The
energy resolution, determined by $\Delta \omega = \frac { \pi}{T}$,
is, in consequence, equal to 0.1~eV.
The time step is
11.025~$\times 10^{-3}$~eV$^{-1}$, and the damping factor used in the Fourier transform
is 0.095~eV.
Troullier-Martins pseudopotentials~\cite{TM}
including non-linear partial
core corrections~\cite{NLCC} for the exchange-correlation interaction between valence and
core electrons, and  an
auxiliary real-space grid~\cite{Daniel}
equivalent to a plane-wave cutoff of 70~Ry
are also  used in this calculation. The basis set includes 13
NAOs per atom: two radial shapes to represent the 3$s$ states plus a
polarization~\cite{Artacho}
$p$ shell with confinement radii $r_s$=$r_p^{Pol.}$=12.2~a.u., and two additional
3$p$ and 3$d$ shells with radii $r_p$=$r_d$=10.0~a.u..

Fig.~1 and Fig.~2 present, respectively, our results for the dipole strength function
and the imaginary part of the linear polarizability of Na$_8$
for energies up to 4~eV. The shape of these curves is in excellent agreement
with both, the calculations of Vasiliev {\it et al.}~\cite{Vasiliev} and,
the experiments of Wang  {\it et al.}.~\cite{Wang}
However, the results appear to be
shifted to higher energies. In fact,
the maximum of the plasmon peak is
obtained at 2.86~eV, which is 0.33~eV higher than the experimentally observed
value.
This shift to higher energies
seems to be related to the extension of the LCAO basis:
using more confined orbitals we get a larger shift.
The integrated dipole strength is equal to 6.97 out of 8, thus fulfilling
87.13~${\%}$ of the sum rule. The partial fulfillment of the sum rule
signifies the incompleteness of
our basis set. The
static linear polarizability $\alpha(0)$ can be obtained from standard (static) calculations
of the induced dipole as a function of the applied field. Using this approach we
obtain a value of 13.2~\AA$^{3}$/atom. An alternative way to calculate $\alpha(0)$
is provided by the formula
\begin{equation}
\alpha(0)=\frac{e^{2} \hbar}{m} \int_{0}^{\infty} \frac {S(\omega) d\omega}{\omega^{2}}
= \frac{2}{\pi} \int_{0}^{\infty} \frac {\Im \alpha(\omega)}{\omega} d\omega,
\label{eq-static}
\end{equation}
from which we obtain a value of 12.5~\AA$^{3}$/atom.
(This result can also be derived from the fact that for the step
perturbation $D(t=0) = \alpha(0) E$.)  The
discrepancy between both estimations is probably related with the
lack of energy resolution of the calculated $\alpha(\omega)$ to
perform the integral in Eq. (\ref{eq-static}) with the required
accuracy. Both results are in reasonable agreement with the values
of 14.6~\AA$^{3}$/atom  and 14.7~\AA$^{3}$/atom, computed by
Vasiliev {\it et al.}~\cite{Vasiliev,Vasiliev2} using the TDLDA
and finite field methods, respectively.

\subsection{Large molecules: C$_{60}$}

The best known Buckyball
C$_{60}$ is a very interesting system with strong electron-electron
interactions due to the confinement. There are quite a
few calculations concerning the optical properties of C$_{60}$ and in
particular its optical response. The main feature of the
optical response of C$_{60}$ is the presence
of two collective excitations (plasmons). The low energy plasmon can be
associated with the ${\pi}$ electrons while the high energy plasmon with both
the ${\sigma}$ and ${\pi}$ electrons, in analogy with the plasmons in graphite.~\cite{Spain,Saito}
The plasmons have been observed by a plethora
of experiments.~\cite{Ajie,Heath,Keller,Hertel,Leach}

The earliest theoretical work~\cite{Bulgac,Rubio,Bulgac2,Alasia}
on C$_{60}$ involved simplifying approximations
for the electron states (tight-binding or spherical averaging) and for the
electron interaction (neglect or RPA-like treatments).
We will compare our results with those of
Westin~{\it et al.}~\cite{Westin} and Yabana and Bertsch~\cite{Yabana},
who used large basis sets and realistic carbon potentials.
Westin~{\it  et al.} used single particle wavefuctions, determined from a
local density approximation (LDA)
calculation, to evaluate the dipole matrix elements which combined with a sum
over states approach yielded the unscreened frequency-dependent linear
polarizability. Screening was included in a RPA-like fashion by introducing an
effective screening parameter. The polarizability calculated in the static
limit was used to evaluate this parameter for the calculation of the dynamic
response. The optical response and the sum rule for the low energy part were in reasonable agreement
with the experiment of Leach~{\it et al.}.~\cite{Leach}
Yabana and Bertsch~\cite{Yabana} used TDLDA, evolving the system in real space
and time, to calculate the dipole strength function of C$_{60}$. Their
calculation also gives reasonable agreement with the experimental data of
Leach~{\it et al.},~\cite{Leach} for
the sum rule of the low energy part although it misses many details of the
structure. This calculation is very similar in quality to ours.

\begin{figure}[]
\centering
\epsfig{file=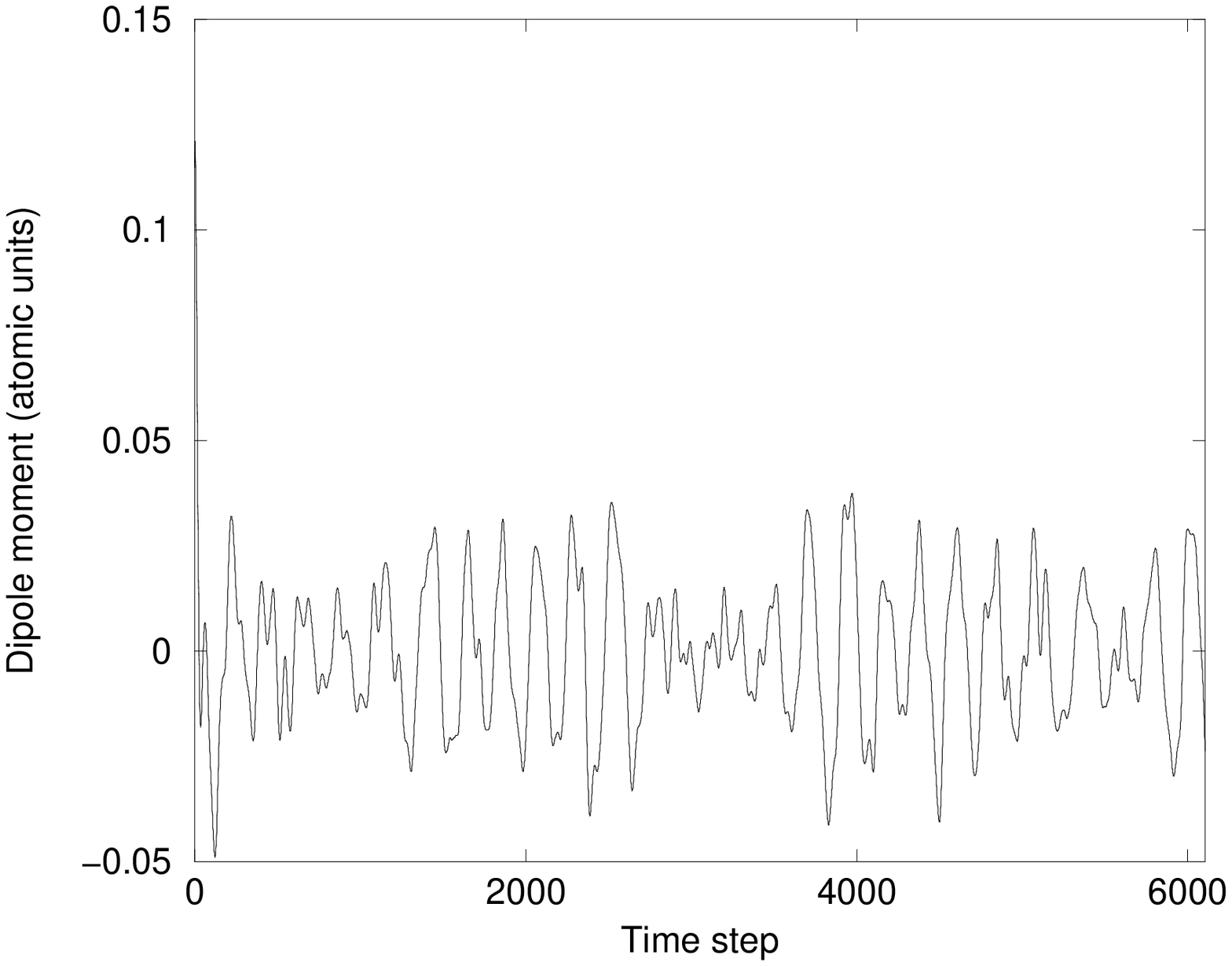, width= 8 cm, angle = 0}
\caption{Dipole moment of C$_{60}$ vs. number of time steps.}
\vspace{0.5 cm}
\centering
\epsfig{file=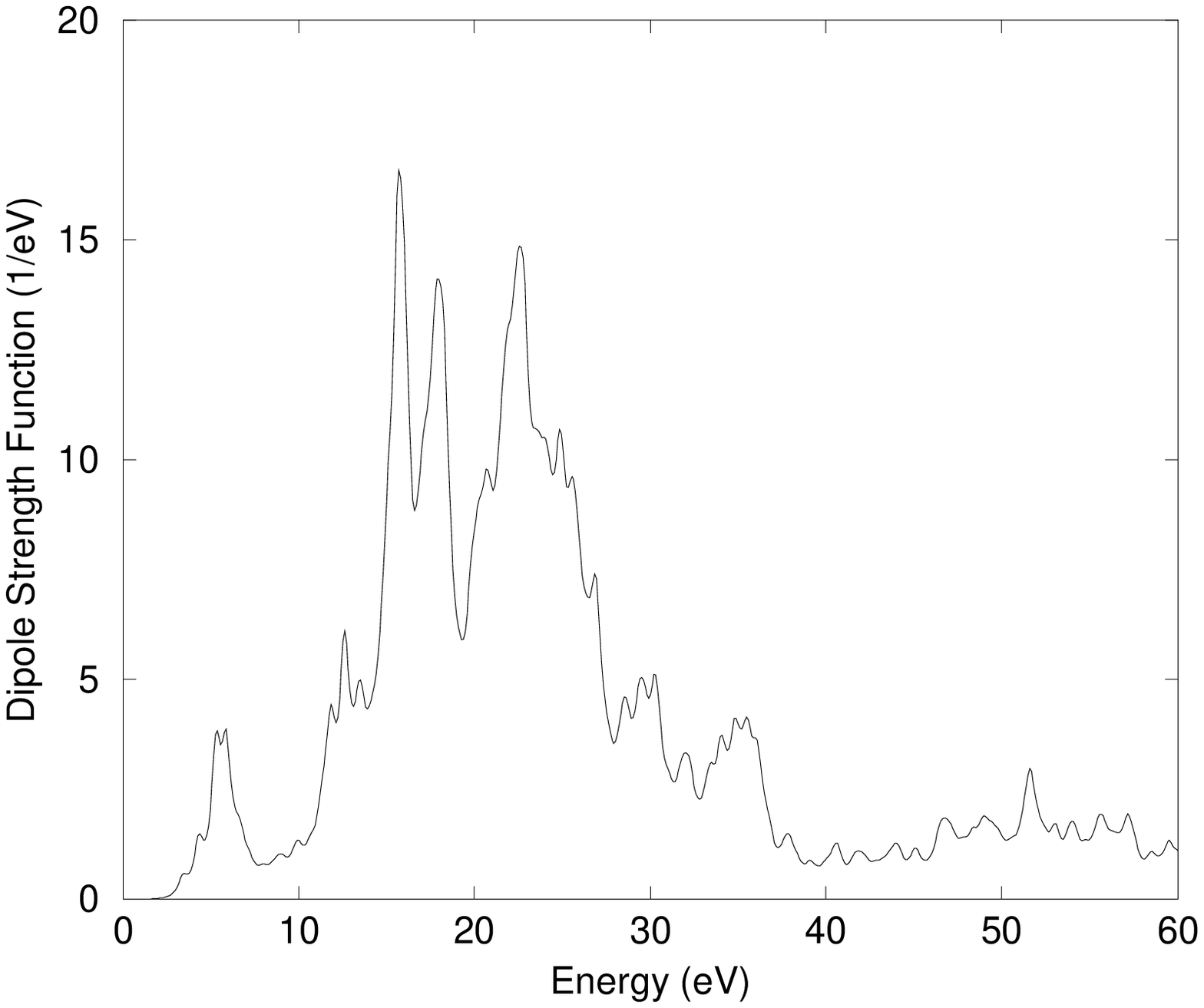, width= 8 cm, angle = 0}
\caption{Dipole strength function of C$_{60}$ vs. energy.}
\vspace{0.5 cm}
\centering
\epsfig{file=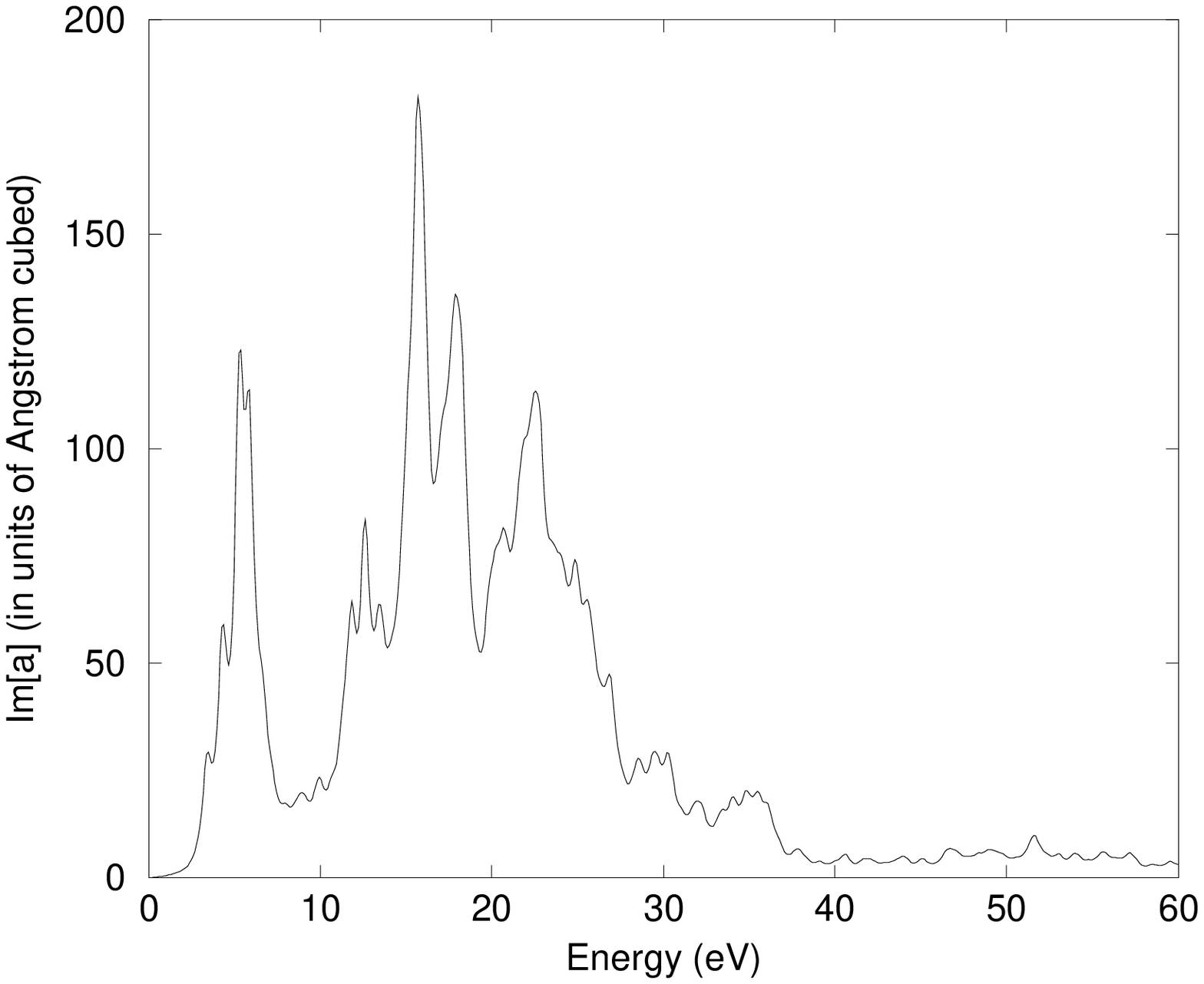, width= 8 cm, angle = 0}
\caption{Imaginary part of the linear polarizability of C$_{60}$
 vs. energy.}
\end{figure}

The total simulation time in our calculation of the polarizability and dipole strength of
the C$_{60}$ molecule is again 31.416~eV$^{-1}$, and the corresponding maximum
energy resolution of 0.1~eV.
The time step however, which is set equal to 5.145 $\times 10^{-3}$~eV$^{-1}$,
is smaller than the one used for Na$_{8}$.
This is
because of the higher frequency range of the response of C$_{60}$. The damping factor used in the
Fourier transform is equal to 0.34 eV in this case.
Troullier-Martins pseudopotentials~\cite{TM}, a double-$\zeta$ polarized basis set, and
a real-space grid
cutoff~\cite{Daniel} of 70~Ry were used in this calculation.
There are 13 NAOs per C atom:
two different radial shapes for the description of the
2$s$ states, another two for the 2$p$,
plus an additional shell of
$d$ orbitals.
The radii of confinement used are $r_s$=5.12~a.u. and $r_p$=$r^{Pol.}_d$=6.25~a.u. (corresponding
to an {\it energy shift}~\cite{Artacho} of 50~meV).
For C$_{60}$, the calculated spectra
show small dependence in these radii, at least as
far as they are not selected to be very stringent.
In Fig.~3, the dipole moment is shown as a function of the time step number.
The dipole strength function obtained
from the time evolution of the dipole moment is shown in Fig.~4 for energies
up to 60~eV. Its main
features are the low energy transitions that come from the ${\pi}$
electrons and the ${\sigma}$ and ${\pi}$ electron transitions in the region of
14-27 eV. In the low energy part of dipole strength function we have peaks at
3.46, 4.35, 5.36, and 5.84 eV, which agree very well with the
ones obtained by the calculations of Westin~{\it et al.}.~\cite{Westin}
By integrating the
dipole strength function over energy we get the sum rule strength. The total sum rule
strength is 223.78 out of 240. Therefore, we satisfy the sum rule up to 93.24~${\%}$.
This reflects the incompleteness of our basis set,
which fails to reproduce some of the excitations in the high energy part of
the spectrum.
The ${\sigma}$ plasmon is broadened,
but this
is a common feature of
all the TDDFT calculations done for C$_{60}$.~\cite{Yabana} In Fig.~5,
the imaginary part of the polarizability is given as  function of energy. By
using Eq. (\ref{eq-static}), the static linear polarizability $\alpha(0)$
is found to be 91.1~\AA$^{3}$,
while our finite field calculations produce a
value of 87.3~\AA$^{3}$. Results for $\alpha(0)$, from very accurate
finite-field calculations using fifteen values of the field, are given in
section \ref{sec-Non}. Both values are higher than the lower limit
estimation of 62.5 \AA$^{3}$ from quantum-mechanical
calculations,~\cite{Fowler}
and in good agreement with the value of 85 \AA$^{3}$ obtained
by Yabana and Bertsch.~\cite{Yabana} They also agree well with the
experimental values of 79.3 \AA$^{3}$ from UV absorption,~\cite{Ren} and
85.2~\AA$^{3}$ from EELS spectra.~\cite{Sohmen,Cohen}

\section{Non-linear polarizabilities}\label{sec-Non}

Because of the non-perturbative nature of our method we are able, for large
values of the applied field, to obtain non-linear polarizabilities. In this
section, we present the calculation of the imaginary part of the integrated
frequency-dependent second order non-linear
polarizability, $\Im \tilde{\gamma}_{step}(\omega)$,
which is related to the response to a step function, for C$_{60}$.
Because  C$_{60}$ is
centrosymmetric the first order non-linear polarizability,
$\beta(\omega)$, and
all other polarizabilities involving
an even number of fields, vanish by symmetry.

The advantage of the explicit time method is that exactly the
same methods can be used to calculate the non-linear response
of the system.  The disadvantage is that (unlike the linear case
where each Fourier component is independent) the non-linear
response depends upon the detailed spectrum of the applied
field.  Here we derive the non-linear response of an electric
field coupled to a C$_{60}$ for the case where the field is
the step function used before.  A different calculation would have
to be done to find the non-linear response to a field with a different
time dependence.

First we give the relation of our calculation to the
general definition of second order non-linear response,
as a function of time, which is~\cite{Flytzanis}
\begin{eqnarray}
\label{eq-nonlinpol}
D^{(3)}(t)= \int_{-\infty}^{t} dt_{1} \int_{-\infty}^{t} dt_{2}
\int_{-\infty}^{t}dt_{3} \\ \nonumber
\gamma(t;t_{1},t_{2},t_{3}) E(t_{1}) E(t_{2})
E(t_{3}).
\label{eq-nonlinpol}
\end{eqnarray}
For the case of a step function perturbation i.e. $E(t) = E \; \theta(-t)$, it
takes the form
\begin{eqnarray}
D^{(3)}(t)= i E^{3} \lim_{\delta_{i} \rightarrow 0^{+}}\label{eq-nonlinp2}
\int \frac{ d\omega_{1} d\omega_{2} d\omega_{3}} {(2 \pi)^{3}} \\
\nonumber \frac{
  e^{-i(\omega_{1}+\omega_{2}+\omega_{3})t}
  \gamma(-\omega_{1}-\omega_{2}-\omega_{3};\omega_{1},\omega_{2},\omega_{3})}
{(\omega_{1}-i\delta_{1})(\omega_{2}-i \delta_{2}) (\omega_{3}-i \delta_{3})}.
\end{eqnarray}
We Fourier transform Eq. (\ref{eq-nonlinp2}) and obtain the second order
non-linear response as a function of frequency
\begin{eqnarray}
D^{(3)}(\omega)= i E^{3} \lim_{\delta_{i} \rightarrow 0^{+}}
\int \frac { d\omega_{2} d\omega_{3}} {(2 \pi)^{2}} \label{eq-nonlinp3}  \\
\nonumber 
\frac{  \gamma(-\omega;\omega - \omega_{2}-\omega_{3},\omega_{2},\omega_{3})}
{(\omega - \omega_{2}-\omega_{3}-i\delta_{1})(\omega_{2}-i \delta_{2})
  (\omega_{3}-i \delta_{3})}.
\end{eqnarray}

The quantity we calculate is the real part of the second order
non-linear response, $\Re D^{(3)}(\omega)$, from which we
can extract the imaginary part of
the integrated second order non-linear polarizability,
$\Im \tilde \gamma_{step} (\omega)$.  Explicit details are
given below and in analogy to Eq.~(\ref{eq-lin0})
 $\Im \tilde \gamma_{step} (\omega)$ is given by
\begin{eqnarray}
\Im \tilde \gamma_{step} (\omega) =  - \omega \lim_{\delta_{i}
\rightarrow 0^{+}}
\int \frac { d\omega_{2} d\omega_{3}} { (2 \pi)^{2}} \label{eq-gammat} \\
\nonumber 
\frac{ \Im  \gamma(-\omega;\omega - \omega_{2}-\omega_{3},\omega_{2},\omega_{3})}
{(\omega - \omega_{2}-\omega_{3}-i\delta_{1})(\omega_{2}-i \delta_{2})
  (\omega_{3}-i \delta_{3})} = \omega \frac{ \Re D^{(3)}(\omega)} { E^{3}}.
\end{eqnarray}
 Just as in Eq.~(\ref{eq-static})
 for the linear term, $\Im \tilde \gamma_{step} (\omega)$
can be related
to the static second order non-linear
polarizability by the expression
\begin{equation}
\frac{2}{\pi}  \int_{0}^{\infty} \frac{d\omega}{\omega} \Im \tilde \gamma_{step} (\omega) = \gamma(0;0,0,0).
\label{eq-rel}
\end{equation}
Eq.~(\ref{eq-rel}) can be trivially  derived by realizing that
$D^{(3)}(t=0) =\gamma(0;0,0,0) E^{3}$ when a step function
perturbation is applied. Alternatively, we can
derive Eq.~(\ref{eq-rel}) directly from Eq.~(\ref{eq-nonlinp3}) by
applying the Kramers-Kroning relations for
$\gamma(-\omega_{1}-\omega_{2}-\omega_{3};\omega_{1},\omega_{2},\omega_{3})$.
In fact, with the help of the Kramers-Kroning relations we can derive
another interesting equality for the first moment
of our integrated response,
\begin{equation}
{1\over 3}\int d\omega \:\: \Im \tilde{\gamma}_{step} (\omega) =
 \int d\omega \:\: \Im \gamma(-\omega;\omega,0,0).
\label{eq-fm}
\end{equation}

For the calculation of $\tilde{\gamma}_{step}(\omega)$ we calculate the response of
the system under two different step function perturbations. In the first
calculation the field used is
equal to $E_{1}=0.10$~V/\AA, and we assume that the response
$D_{1}(\omega)$ is linear with
respect to the field. This assumption is true since for $\omega = 0$ the
non-linear terms contribute to the response only 4.82$\times$10$^{-3} \%$.
The contribution is of the same order of magnitude for $\omega \neq
0$. In the second calculation the field is equal to
$E_{2}=1.00$ V/\AA, and we assume that the response $D_{2}(\omega)$
consists of the linear response and the second order non-linear response
$D_{2}^{(3)}(\omega)$. The values of the field are in the same range as
those used by Westin {\it et al.}~\cite{Westin} for the determination of the
static second order non-linear polarizability.
Using Eq.~(\ref{eq-lin0}), we have that
\begin{equation}
D_{1}(\omega)=\alpha(\omega) E_{1}(\omega)= \alpha(\omega)  \frac{ E_{1}}{i
\omega} \label{eq-lin}
\end{equation}
and
\begin{equation}
D_{2}(\omega) - \alpha(\omega) E_{2}(\omega)= D_{2}^{(3)}(\omega). \label{eq-nonlin}
\end{equation}
From Eq. (\ref{eq-gammat}) it follows that 
\begin{equation}
D_{2}^{(3)}(\omega)= \frac { \tilde{\gamma}_{step}(\omega) } { i \omega} 
E_{2}^{3}, \label{eq-ndelta}
\end{equation}
and from Eqs. (\ref{eq-lin}), (\ref{eq-nonlin}), and (\ref{eq-ndelta})
we obtain
\begin{equation}
\tilde{\gamma}_{step}(\omega)= \frac{i \omega} {E_{2}^{3}} \left( D_{2}(\omega) -
\frac{E_{2}}{E_{1}} D_{1}(\omega) \right).
\end{equation}
Our calculation for $\tilde{\gamma}_{step}(\omega)$ is quite straightforward
in contrast to the perturbative method where it becomes computationally very
demanding.

\begin{figure}[t]
\centering
\epsfig{file=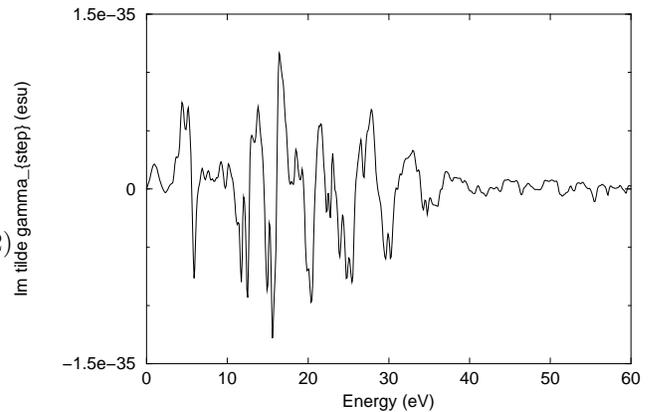, width= 8 cm, angle = 0}
\caption{Imaginary part of the integrated second order non-linear
polarizability , $\Im \tilde{\gamma}_{step}(\omega)$, of C$_{60}$ vs. energy.}
\end{figure}

In Fig. 6, we present the results, up to 60~eV, for
$\Im \tilde{\gamma}_{step}(\omega)$, where
$\Im \tilde{\gamma}_{step}(\omega)$ is given
by Eq.~(\ref{eq-gammat}).
As expected, $\Im \tilde{\gamma}_{step}(\omega)$ has both
positive and negative values.
The reason why $\Im \tilde{\gamma}_{step}(\omega)$ does not
vanish below some finite frequency (as does the
linear response) is because the second-order non-linear
term represents many processes of both absorption
and emission of photons and the C$_{60}$ molecule
can couple to a continuum of modes that extends to zero
frequency.  This can also be seen in the integral
expression Eq.~(\ref{eq-gammat}).

Similarly to the case of the linear polarizability, we can obtain
an estimation of the magnitude of $\gamma(0;0,0,0)$ from static
self-consistent calculations performed with finite fields.
This value can be contrasted to similar calculations in the
literature, providing an estimation of the uncertainty of our
calculations of the non-linear terms, and
an internal consistency test for our calculation of the
integrated frequency-dependent second order non-linear polarizability.
We have followed here a procedure similar to that used in
Ref.~\onlinecite{Quong}, performing LDA calculations of the total
energy and electric dipole of the C$_{60}$
molecule for fifteen different values of
an external static electric
field E ranging from 0.003~V/\AA\
to 2.0~V/\AA. The results of the total
energy were then
fitted using the expression W$_{tot}$
=W$_{0}$-$\frac{1} {2} \alpha$E$^2$-$\frac{1}{4} \gamma$E$^4$, where $\alpha$ is the
linear polarizability and $\gamma$ the second order non-linear
polarizability. The values obtained for $\alpha$ and $\gamma$ are,
respectively, 85.3~\AA$^3$ and 3.54$\times$10$^{-36}$~esu.
The results for the dipole moment were fitted to the expression
D=$\alpha$E+$\gamma$E$^3$.
The corresponding values obtained are 85.3~\AA$^3$ and
3.71$\times$10$^{-36}$~esu. These values of $\gamma(0;0,0,0)$ are in excellent
agreement with the value of 3.72$\times$10$^{-36}$ esu obtained using
Eq. (\ref{eq-rel}), thus confirming the validity of our calculation for
$\tilde{\gamma}_{step} (\omega)$. All of
the values obtained by our calculations are
smaller than the upper bound of 3.7$\times$10$^{-35}$ esu proposed by the experiments
of Geng and Wright.~\cite{Wright}
Also, our results for $\gamma(0;0,0,0)$ are in reasonably good agreement with
other LDA calculations in the literature.
Quong {\it et al.}~\cite{Quong} reported values
of 82.7~\AA$^3$ and 7.0$\times$10$^{-36}$~esu, for $\alpha$ and $\gamma$,
respectively, using an all-electron method with a Gaussian expansion as
a basis set. Van Gisbergen {\it et al.}~\cite{TDLDAgamma} reported
very similar values,
82.5~\AA$^3$ and 7.3$\times$10$^{-36}$~esu, using a computational
scheme based on
a frozen-core approximation, and a basis set of Slater functions.
It is interesting to note that these results are much smaller
than those obtained using
simplified tight-binding models within an independent
electron picture, where the effects of screening are neglected. In such calculations the values of
$\gamma(0;0,0,0)$ obtained are of the order
of $~$200$\times$10$^{-36}$~esu.~\cite{Shuai,Harigaya,Dong}

\section{Conclusion}\label{sec-Concl}

We presented a method for the calculation of the optical response of atoms and
clusters. The main features of the method are the description of the
wavefunctions in terms of an efficient local orbital (LCAO) basis and the
explicit evolution of the system in time. This approach is designed for large
clusters and in fact it gives excellent results for C$_{60}$. It is also
shown to work remarkably well even for
systems for which the LCAO basis is very small, such as Na$_{8}$.
Our approach has the desirable features that only occupied states are
needed and that all the most computationally intensive operations are
essentially the same as those used to calculate the ground state
properties. In addition, non-linear effects can be included in a straightforward
way, and we have shown how to calculate the second order non-linear response
for C$_{60}$.

\begin{acknowledgments}

We would like to thank Prof. L. Cooper for the useful
discussions, and Dr. I. Vasiliev for reading the manuscript.
This material is based upon work supported by the U.S. Department of
Energy, Division of Materials Sciences under Award No. DEFG02-ER9645439,
through the Frederick Seitz Materials Research Laboratory.

\end{acknowledgments}

\appendix

\section*{Alternative approaches for the solution
of the time-dependent Kohn-Sham equation}

Other ways to approximate the exponential in Eq. (\ref{eq-appr}) include the
expansion of the exponential in a series of Chebyshev polynomials~\cite{Tal}
\begin{eqnarray}
exp\left(-i S^{-1}H \Delta t \right)
\simeq exp\left(-i(\Delta E/2+E_{min})\Delta t \right)\times \\
\sum_{n=0}^{N} a_{n}\left(\frac {\Delta E~t}{2}\right) \nonumber
\phi_{n}(H_{norm})
\end{eqnarray}
where $\phi_{n}$ are the Chebyshev polynomials,
and the expansion coefficients $a_n(x)$ can be shown to be analogous
to Bessel functions of
the first kind of order n.
$H_{norm}$ is a normalized Hamiltonian
$2(S^{-1}H-E_{av})/\Delta E$, where $E_{av}$=($E_{max}+E_{min}$)/2,
$\Delta E$=$E_{max}-E_{min}$. $E_{max}$ and $E_{min}$ are, respectively,
the maximum and minimum eigenvalues of $S^{-1}H$.
The Chebyshev polynomials are chosen
because their error decreases exponentially when N is
large enough, due to the
uniform character of the Chebyshev expansion.~\cite{Tal}
Time reversal is built into the expansion
coefficients, but the method does not effectively
conserve norm or energy. While it works remarkably well for
time-independent Hamiltonians,~\cite{Lef}
for time-dependent Hamiltonians the method
becomes inefficient as the
Chebyshev polynomials of the Hamiltonian have
to be recalculated for each time step.

Another approximation for the exponential in Eq. (\ref{eq-appr}) is performed
by using the split-operator method introduced by Feit~{\it et al.}~\cite{Feit}
According to this
method, the exponential which contains the Hamiltonian operator can be split as
\begin{eqnarray}
exp[-i(T+V)\Delta t] \simeq    \label{eq-split} \\
exp(-i \frac {1} {2} T \Delta t)&  \; exp(-i V \Delta t)
\; exp(-i \frac {1} {2} T \Delta t). \nonumber
\end{eqnarray}
This method was later generalized by Suzuki~\cite{Suzuki1} for an arbitrary number
of operators, providing higher order expansions (formula (\ref{eq-split}) corresponds
to second order in $\Delta t$), and a rigorous extension
of the method to time-dependent Hamiltonians.~\cite{Suzuki2}
The split operator method takes advantage of the fact that it is very convenient
to treat operators in their diagonal representations.
For example, it is trivial to apply the kinetic energy operator to a wave-function
in Fourier space, while the effect of a local potential is more easily calculated
in real space.
This method is, in principle, unconditionally stable and norm conserving. On the
other hand, it does not conserve energy and it can only be used to Hamiltonian
operators which can be split into two non-commuting parts with a simple
transformation between them. Therefore, the method is
very well suited for plane wave or real
space methods, where efficient fast Fourier transform algorithms provide
an {\it exact} transformation between {\it finite} plane-wave expansions
and real
space grids.
In other words,
they span
exactly the same subspace of functions, and it is possible
to switch between representations where the kinetic
energy and the potential are diagonal within a given subspace.
For an LCAO basis this is not possible. If we
take an arbitrary wavefunction expanded in an orbital basis
$\psi({\bf r})= \sum_\nu c^\nu \phi_\nu ({\bf r})$, the result of
applying the operator $exp(-iVt)$
using a grid of points in real-space $f({\bf r}_i)=exp(-iV({\bf r}_i) t) \psi({\bf r}_i)$
will be, in general,
not representable using the same local basis, i.e.
some of the resulting function has been spilled from
the subspace spanned by the local basis.

An alternative way to solve Eq. (\ref{eq-LCAO}) is by using the second-order
differencing (SOD) method introduced by Askar and Cakmak.~\cite{Askar}
In SOD the symmetric relation is used
\begin{equation}
c(t+ \Delta t) - c(t - \Delta t) = ( e^{-i S^{-1} H(t) \Delta t} -  e^{ i S^{-1}
H(t) \Delta t}) c(t)
\end{equation}
and by expanding the exponential in Taylor series, the second-order propagation
scheme is obtained
\begin{equation}
c(t+ \Delta t) \simeq  c(t - \Delta t) - 2 i \Delta t S^{-1} H(t) c(t).
\end{equation}
By construction SOD obeys the time-reversal symmetry. SOD is stable, and
norm and energy are approximately conserved, only
if the Hamiltonian is
Hermitian. In spite of its simplicity, an important
drawback of this method for its application to large systems is that
it is necessary to store the wavefunctions at two different time steps,
which may require a large amount of memory.

Finally, another method for solving Eq. (\ref{eq-LCAO}) is the short
iterative Lanczos.~\cite{Lef,Park}
It is very convenient for calculations that involve
very big Hamiltonians, especially when they are time independent.
The
Hamiltonian is projected to a subspace of smaller
dimensionality and takes a tridiagonal form that makes easy to perform
calculations. The Lanczos recurrence creates a set of orthogonal
polynomials which constitute a finite polynomial approximation of the
operator. An interesting feature of the method is its dependence on both the
operator and the initial vector. The Lanczos recurrence relation is
\begin{equation}
(S^{-1}H)\;{\bf q_{j}}=\beta_{j-1}{\bf q_{j-1}}+
\alpha_{j}{\bf q_{j}}+\beta_{j}{\bf q_{j+1}}.
\end{equation}
The coefficients are $\alpha_{j}=({\bf q_{j}},(S^{-1}H)\;{\bf q_{j}})$ and
$\beta_{j-1}=({\bf q_{j-1}},(S^{-1}H)\;{\bf q_{j})}$, where
$({\bf a},{\bf b})={\bf a}^{\dagger}S{\bf b}$ is
the usual complex
inner product, and
$({\bf q_{i}},{\bf q_{j}})=\delta_{ij}$.
Matrices H and S have
$N_b \times N_b$ dimension and the vectors ${\bf q_j}$ have $N_b$
components, $N_b$ the total number of basis functions.
For very large $N_b$, where the method is particularly useful,
the inverse $S^{-1}$ is not directly calculated but an
iterative method is used instead.~\cite{Park}
The recurrence relation
is initiated by setting
\begin{equation}
{\bf q_{0}}={\bf c}(0)
\end{equation}
and
\begin{equation}
(S^{-1}H)\;{\bf q_{0}}=\alpha_{0}{\bf q_{0}}+\beta_{0}{\bf q_{1}}.
\end{equation}
After $P$ iterations, the projected operator
$(S^{-1}H)_{P}=({\bf q_{j}},(S^{-1}H)\;{\bf q_{i}})$
is a
$P\times P$ matrix with a tridiagonal form that can be very easily
diagonalized.
The solution of Eq. (\ref{eq-LCAO}) becomes
\begin{equation}
{\bf c}(\Delta t)= Z^{\dagger} e^{-i D_{P} \Delta t} Z {\bf c}(0),
\label{tridiagonal}
\end{equation}
where $Z$ is the $N_b \times P$ transformation
matrix that diagonalizes $(S^{-1}H)_{P}$ and $D_{P}$ is the
diagonalized matrix.
For an arbitrary initial condition, the accuracy of the
time evolution achieved with Eq. (\ref{tridiagonal})
is equivalent to a $P$ order expansion of the evolution operator
\begin{equation}
{\bf c}(\Delta t) = \sum_{k=0}^{P-1}
{\frac {-it^k} {k!} } (S^{-1}H)^k \; {\bf c}(0),
\end{equation}
and, consequently, the energy is only approximately conserved.
However, the propagator in Eq. (\ref{tridiagonal})
is unitary, and the normalization condition
is strictly
preserved. For time independent Hamiltonians, Eq. (\ref{tridiagonal})
can be used to evolve the wavefunction for a time interval
$\tau$
that depends on $P$. For times larger than $\tau$,
the time evolution predicted by a $P$ order expansion
becomes inaccurate, and it is necessary to recalculate
propagator using ${\bf c}(\tau)$ as the starting point for
the recurrence procedure.
The method is, therefore, very well suited to follow
the evolution of wavefunctions described by a large
basis set during short periods of time ($<\tau$).

For the purpose of the calculation of response
functions,
we believe that the method
adopted in
this paper, based on the Crank-Nicholson operator,
is superior to the short iterative Lanczos,
at least for moderate basis sizes.
There are several reasons for this:
{\it i)} We need to
evolve the wavefunctions
for long times.
{\it ii)} The Hamiltonian is time dependent. This implies that the
recurrence cycle for the calculation of the approximated
time evolution operator has to be
repeated for each time step. {\it iii)}
In our case, we have to
evolve all the occupied states. The propagator
in Eq.(\ref{tridiagonal}), however, depends on the initial state.
Therefore,
it is necessary to develop some generalization of
the scheme presented. For example,
it might be also possible to construct
an approximate time evolution operator
starting from some weighted linear
combination of the occupied states,
and projecting it into a subspace of
dimension $P>N_{occ}$, where $N_{occ}$
is the number of occupied wavefunctions.


\begin{references}
\bibitem{Efield} Our calculations are performed using a supercell
geometry. Therefore, we are talking
here
about the ground state of
a bounded system in the presence of a periodic triangular potential,
with slope $E$. This provides a meaningful model for a cluster in the presence of a
finite electric field only if $|E|<|E_c|$, where the critical slope
depends on the system under consideration, and
on the lateral size of the simulation
cell $L$. In particular,
$E_c \rightarrow 0$ when $L \rightarrow \infty$.
In our calculations of the linear response we use
small electric fields,
such that
$|E| << |E_c|$.
\end{references}
\end{document}